\documentclass[a4paper, amsfonts, amssymb, amsmath, reprint, showkeys, nofootinbib, twoside,notitlepage,onecolumn]{revtex4-2}

\usepackage{amsmath,amstext}
\usepackage[T1]{fontenc}
\usepackage{amssymb}
\input{epsf}
\usepackage{graphicx}
\usepackage{ae,aecompl}

\usepackage{hyperref}
\usepackage{amsmath}
\usepackage{amssymb}
\usepackage{mathtools}
\usepackage{bm}
\usepackage{cleveref}
\usepackage{tensor}
\usepackage{braket}
\usepackage{enumitem}
\usepackage{mhchem}
\usepackage{cancel}
\usepackage{amsthm}
\usepackage{upgreek}
\usepackage{mathrsfs}

\usepackage{color}

\newcommand{\spc}{\quad \quad \quad}

\newcommand{\beqy}{\begin{eqnarray}}
\newcommand{\eeqy}{\end{eqnarray}}

\def\be{\begin{equation}}
\def\ee{\end{equation}}
\def\beq{\begin{eqnarray}}
\def\eeq{\end{eqnarray}}

\theoremstyle{definition}

\theoremstyle{theorem}
\newtheorem{theorem}{Theorem}

\theoremstyle{corollary}

\begin{document}
\title{Thermodynamic stability of superflows in General Relativity and Newtonian gravity}
\author{ L.~Gavassino\footnote{\textbf{email:} lorenzo.gavassino@gmail.com}}
\affiliation{
Department of Mathematics, Vanderbilt University, Nashville, TN, USA
}

\begin{abstract}
Landau's criterion for superfluidity is a special case of a broader principle: A moving fluid cannot be stopped by frictional forces if its state of motion is a local minimum of the grand potential. We employ this general thermodynamic criterion to derive a set of inequalities that any superfluid mixture (with an arbitrary number of order parameters) must satisfy for a certain state of motion to be long-lived and unimpeded by friction. These macroscopic constraints complement Landau's original criterion, in that they hold at all temperatures, and remain valid even for gapless superfluids. They are only necessary conditions for the existence of a frictionless hydrodynamic motion, since they presuppose the validity of a fluid description, but they provide sufficient conditions for stability against stochastic hydrodynamic fluctuations. We first formulate our analysis within General Relativity (with neutron star applications in mind), and then we take the Newtonian limit.
\end{abstract} 

\maketitle

\section{Introduction}
\vspace{-0.4cm}

How can superfluids undergo perpetual frictionless flow? The standard answer, due to Landau \cite[\S 22]{landau9}, goes as follows: Fix a flow, and compute the energy gain associated with reducing its momentum via individual quantum excitations. A superfluid is a substance for which such gain is negative (i.e. it costs energy to slow it down). This argument, while insightful, is not the whole story, since it applies strictly at zero temperature \cite[\S 13.4]{huang_book}, far below the critical temperature at which superfluidity emerges. Moreover, recent work suggests that certain fermionic superfluids (so-called \textit{gapless} superfluids \cite{AllardChamel2023PartI,AllardChamel2023PartII,AllardChamel2024EPJA,AllardChamel2024PRL,AllardChamel2025}) can violate Landau's criterion, and still do not experience dissipation or breakdown. These observations force one to provide a more refined explanation of superfluidity, which is summarised below.

Let us start with a well-known result. When the laws of thermodynamics are applied to a body in contact with an external medium, one arrives at the following irreversibility statement \cite[\S 20]{landau5}:

\noindent ``\textit{Consider an isolated bipartite system, where a body exchanges energy and particles with an environment at rest. Assume that the internal state of the environment is always an equilibrium state, with constant temperature $T_\star$ and chemical potential $\mu_\star$. Then, the grand potential functional $\Omega=U-T_\star S-\mu_\star N$ cannot undergo macroscopic increase, where $U$, $S$, and $N$ are the body's energy (in the environment frame), entropy, and particle number.}''\footnote{\label{foononanan1ona}\textit{Proof}: Since the total system ``body+environment'' is isolated, a transformation is thermodynamically allowed only if it fulfills the following constraints: $\Delta U+\Delta U_E=0$ (energy conservation), $\Delta S+\Delta S_E\geq 0$ (second law), and $\Delta N+\Delta N_E=0$ (particle conservation), where the subscript ``$E$'' refers to the environment. Combining this with the identity $\Delta U_E=T_\star \Delta S_E+\mu_\star \Delta N_E$, which follows from our assumptions on the state of the environment, we immediately obtain $\Delta(U-T_\star S-\mu_\star N)\leq 0$.\\
\textit{Remark}: Note that, if the body is out of equilibrium, $\Omega$ is not the usual grand potential $U-TS-\mu N$ we encounter in thermodynamics, because $T_\star$ and $\mu_\star$ are properties \textit{of the environment}, and not of the fluid itself. To avoid confusion, some sources refer to $\Omega$ as the ``exergy'' (i.e. extractible energy \cite{BecattiniExergy2023}). In the case of fluid systems, such a distinction is superfluous, since $T$ and $\mu$ are generally nonuniform out of equilibrium, causing the expression $U - T S - \mu N$ to be ill-defined, thereby making $U-T_\star S-\mu_\star N$ the ``only game in town''.}\\
The general law stated above is particularly valuable, as it enables us to determine the state in which a body spends most of its time. In fact, as we let the system evolve, spontaneous irreversible processes and random fluctuations tend to drive $\Omega$ to progressively lower values, until a local minimum is reached. Once there, the system becomes effectively ``trapped'', since any further evolution would require an increase in $\Omega$, which is prohibited, provided that the grand potential barrier separating the given minimum from lower minima is macroscopically large (i.e., diverges as $N \to \infty$). Consequently, if we do not force it from outside, the body will remain in the first local minimum of $\Omega$ it encounters.

\begin{figure}
    \centering
\includegraphics[width=0.30\linewidth]{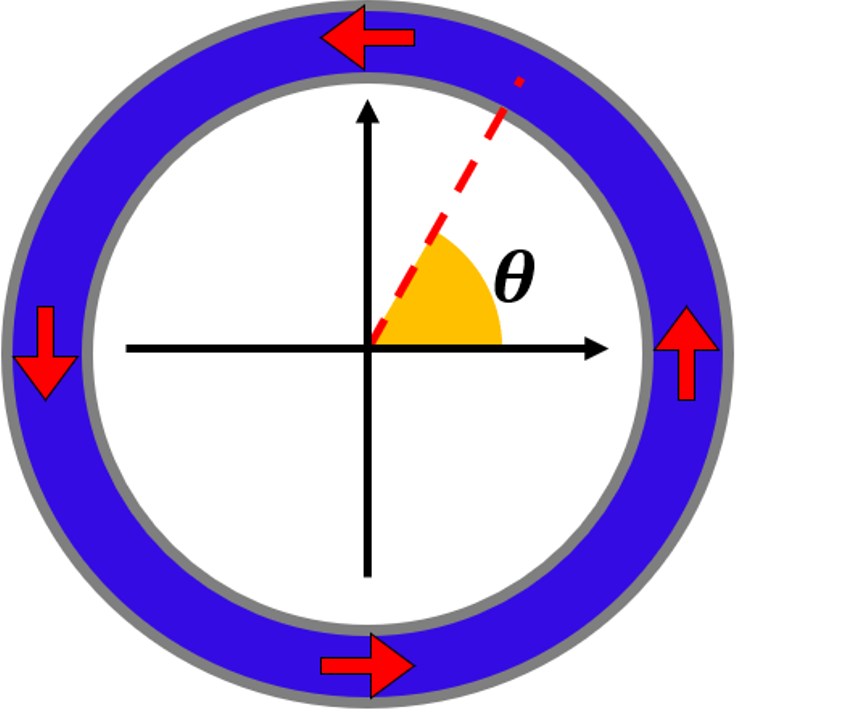}
\includegraphics[width=0.45\linewidth]{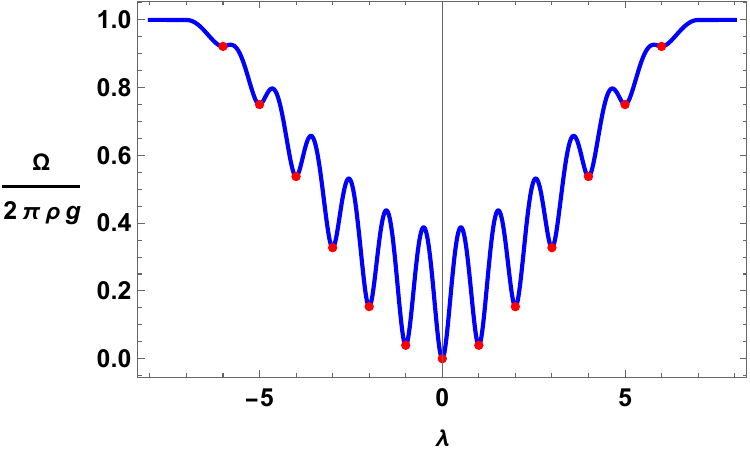}
\caption{Thermodynamic explanation for the long lifetime of superflows. Left panel: A simple experimental setup, where a superfluid is confined in a circular ring-shaped pipe (with negligible thickness). Right panel: In this experimental setup, the grand potential \eqref{grandone} admits a discrete list of local minima $\psi_z$, given by \eqref{minimina}, corresponding to different rotation rates with quantized angular velocity. In the plot, we graph the grand potential $\Omega(\lambda)$ (with $g=\rho=5$) along a continuous one-parameter family of states $\psi(\lambda)=\sum_z c_z(\lambda)\psi_z$, with $c_z =(1-|\lambda{-}z|)\Theta(1-|\lambda{-}z|)$. This family is constructed so that the state $\lambda=z$ has order parameter $\psi_z$, and it thus matches a local minimum (red dots). Mathematically, the grand potential barriers separating the minima exist because changing the winding number of the phase of $\psi$ requires $|\psi|$ to vanish at some location. This entails lifting the order parameter to the top of the Mexican hat potential.
}
    \label{fig:1}
\end{figure}

In most gases and liquids, $\Omega$ has a single minimum (the equilibrium state), which is absolute, and corresponds to a state where the fluid is at rest relative to the environment. From this, one concludes that regular fluids admit no long-lived currents, since frictional forces with the environment impede any state of motion. However, the situation may change if a fluid possesses a complex order parameter $\psi$, arising from a broken $U(1)$ symmetry. To see why, imagine a scenario where the fluid of interest is confined to circulate inside a thin circular pipe (figure \ref{fig:1}, left panel), and postulate a simple Ginzburg-Landau-type grand potential with a Mexican-hat shape:
\vspace{-0.2cm}
\begin{equation}\label{grandone}
\Omega = \int_0^{2\pi} \left[\partial_\theta \psi^* \, \partial_\theta \psi +g (\psi^* \psi -\rho)^2\right] d\theta \spc (\text{with }0\equiv 2\pi) \, ,   
\end{equation}
where $\theta$ is the polar angle around the pipe's barycenter, while $g$ and $\rho$ are two positive constants. To find the local minima of $\Omega$, we consider a one-parameter family of states $\psi(\lambda)$, where $\lambda\,{=}\,0$ is the candidate minimum, and we require that $\dot{\Omega}(0)\,{=}\,0$ and $\ddot{\Omega}(0)\,{\geq}\, 0$ for any choice of $\dot{\psi}(0)$ and $\ddot{\psi}(0)$ (with $\dot{f}=df/d\lambda$), producing the equilibrium conditions
\vspace{-0.2cm}
\begin{equation}
\begin{split}
\dot{\Omega}(0)={}& \int_0^{2\pi} \left[- \dot{\psi}^* \, \partial^2_\theta \psi-\partial_\theta^2 \psi^* \,  \dot{\psi} +2g (\psi^* \psi -\rho)(\dot{\psi}^* \psi+\psi^* \dot{\psi})\right] d\theta =0 \, , \\
\ddot{\Omega}(0)={}& \int_0^{2\pi} \left[2\partial_\theta \dot{\psi}^* \, \partial_\theta \dot{\psi} - \ddot{\psi}^* \, \partial^2_\theta \psi-\partial_\theta^2 \psi^* \,  \ddot{\psi} +2g (\psi^* \psi -\rho)(\ddot{\psi}^* \psi+\psi^* \ddot{\psi}+2\dot{\psi}^* \dot{\psi})+2g(\dot{\psi}^* \psi+\psi^* \dot{\psi})^2\right] d\theta \geq 0 \, .\\
\end{split}
\end{equation}
\newpage

\noindent The condition that the first derivative vanishes leads to the differential equation $\partial^2_\theta \psi = 2g(\psi^* \psi - \rho)\psi$. Combined with the periodic boundary condition $\psi(0) = \psi(2\pi)$, this equation admits the following discrete set of solutions:
\vspace{-0.1cm}
\begin{equation}\label{minimina}
\psi_z(\theta)=\sqrt{\rho-\dfrac{z^2}{2g}} \, e^{iz\theta}  \spc \text{for }z\in \mathbb{Z}\cap [-\sqrt{2\rho g},\sqrt{2\rho g}] \, .
\end{equation}
To determine whether these are actual minima (see figure \ref{fig:1}, right panel), we plug \eqref{minimina} into the formula for $\ddot{\Omega}(0)$. Introducing, for convenience, the rotated field
$e^{-iz\theta}\dot{\psi}\equiv h=h_R +ih_I$ (with $h_R,h_I \in \mathbb{R}$), we obtain
\vspace{-0.1cm}
\begin{equation}
\begin{split}
\ddot{\Omega}(0)={}& \int_0^{2\pi} \left[2\partial_\theta \dot{\psi}^* \, \partial_\theta \dot{\psi} -2z^2 \dot{\psi}^* \dot{\psi}+(2\rho g{-}z^2)(e^{iz\theta}\dot{\psi}^*+e^{-iz\theta} \dot{\psi})^2\right] d\theta  \\
={}& \int_0^{2\pi} \left[2\partial_\theta h^* \partial_\theta h+2iz(h\partial_\theta h^*-h^* \partial_\theta h) +(2\rho g{-}z^2)(h^* {+}h)^2\right] d\theta \\
={}& 2\int_0^{2\pi} \left[(\partial_\theta h_R)^2 +
\begin{pmatrix}
h_R & \partial_\theta h_I \\
\end{pmatrix}
\begin{bmatrix}
2(2\rho g -z^2) & 2z \\
2z & 1 \\
\end{bmatrix}
\begin{pmatrix}
h_R \\
\partial_\theta h_I
\end{pmatrix}
\right] d\theta  \, ,\\
\end{split}
\end{equation}
which is guaranteed to be positive definite (except for $h=i \times \text{``real const''}$, for which $\ddot{\Omega}(0)=0$ due to $U(1)$ symmetry) if the determinant of the $2\times 2$ matrix is positive, namely if $3z^2<2\rho g$.

In summary: If $g \rho$ is large enough, the order parameter admits a discrete list of long-lived configurations \eqref{minimina}, where the phase is non-uniform, and winds around the circumference of the pipe $z \in \mathbb{Z}$ times (with $|z|$ not too large). Adding the assumption (grounded on quantum mechanical reasoning) that the velocity of the fluid is proportional to the gradient of the phase of $\psi$, we conclude that \eqref{minimina} describes a fluid undergoing a perpetual frictionless circular flow, with quantized angular velocity $\omega_z \propto z$. This qualifies the fluid as a \textit{superfluid}, and the perpetual flow as a \textit{superflow}.

This article aims to extend the above calculation to more realistic scenarios. Specifically, we will examine a generic mixture of normal and superfluid components, with an arbitrary equation of state, subject to an arbitrary background gravitational field, and confined within a container of arbitrary shape. {\color{black}Our task will be to identify the geometric and thermodynamic conditions under which a given hydrodynamic flow is a local minimum of $\Omega$, and is thus long-lived (i.e. its lifetime diverges in the thermodynamic limit). We will carry out the calculation in General Relativity, where the four-dimensional index notation streamlines the expressions, and only afterward will we take the Newtonian limit. In this way, our results are equally applicable to superfluid helium and to neutron star interiors (superfluidity plays a crucial role in neutron stars, causing pulsar glitches and modifying the cooling rate \cite{andersson2007review,Page2014}). To the best of our knowledge, such a calculation has never been performed before, except in a few highly specialized settings \cite{Andreev2004,Haber:2015exa,Gouteraux:2022qix,Arean:2023nnn,Hoult:2024cyx}. 

The paper is structured as follows. Section \ref{section2} reviews the relativistic hydrodynamic framework for superfluid mixtures. Section \ref{stabiliycrituzzione} derives the conditions under which a superfluid state minimizes the grand potential $\Omega$, showing that local minima can exist also with non-vanishing superfluid motion relative to the environment. Section \ref{RelativisticSingleComponent} applies these results to a single-component superfluid, and Section \ref{nonrellimitsect} discusses the Newtonian limit.}

\textit{Assumptions and conventions}: Throughout the article, we adopt the metric signature $(-,+,+,+)$, and work in natural units, with $c=\hbar=k_B=1$. The spacetime metric (and thus the gravitational field) is treated as an externally fixed background. The superfluid substance is modeled hydrodynamically, which means that we can only minimize $\Omega$ within the manifold of states that are locally thermal. 

\vspace{-0.2cm}
\section{Multicomponent superfluids in general relativity}\label{section2}
\vspace{-0.2cm}

In this section, we briefly review the hydrodynamic framework that is currently used to model superfluid mixtures in General Relativity. The interested reader can see, e.g., \cite{Carter_starting_point,carter92,cool1995,langlois98,Son2001,GusakovAndersson2006,Gusakov2007,Gusakov:2016eom,Herzog2009,Jensen:2012jh,GavassinoKhalatnikov2022,GavassinoStabilityCarter2022} for a variety of alternative derivations, all of which lead to the same theory \cite{Termo}, which is the relativistic analog of the \citet{AndreevBash1975} ``$N-$fluid model'' of superfluid solutions.

\vspace{-0.2cm}
\subsection{The macroscopic degrees of freedom of a superfluid mixture}
\vspace{-0.2cm}

What are the relevant fields of superfluid hydrodynamics? First of all, every superfluid system has a temperature $T(x^\alpha)$ and a normal velocity $u^\nu(x^\alpha)$. These parameters define the local statistical ensemble of the excitations. Specifically, fixed an event $\mathscr{P}$ in spacetime, the density matrix $\hat{\sigma}$ of the excitations in a neighborhood of $\mathscr{P}$ is \cite{cool1995,GavassinoKhalatnikov2022}
\begin{equation}\label{densitymatrix}
\hat{\sigma}\propto e^{\frac{\hat{\mathcal{P}}^\nu u_\nu(\mathscr{P})}{T(\mathscr{P})}} \, ,
\end{equation}
where $\hat{\mathcal{P}}^\nu$ the four-momentum operator of the excitations in a box surrounding $\mathscr{P}.$\footnote{\label{footnoteonunu}Note that a notion of normal velocity exists also at zero temperature. In fact, in this limit, the density matrix is just the ground state of the Hamiltonian $\hat{H}=- \hat{\mathcal{P}}^\nu u_\nu$, which means that the excitations minimize the energy \textit{in the normal rest frame}. Indeed, Landau's criterion for superfluidity \cite[\S 22]{landau9} may be stated as follows: In a single-component superfluid at $T=0$, with order parameter $\psi=|\psi| e^{i\varphi}$, excitations that lower $|\psi|$ spontaneously ``pop up'' \textit{if and only if} the ground state of the Hamiltonian in the normal frame, $- \hat{\mathcal{P}}^\nu u_\nu$, does not coincide with the ground state of the Hamiltonian in the superfluid frame, $- \Hat{\mathcal{P}}^\nu \nabla_\nu \varphi /|\nabla \varphi|$.} 

Besides $T$, we need some more thermodynamic variables. In particular, we need a list of scalars $\mu^A(x^\alpha)$, representing the chemical potentials of all the ``normal species'', i.e. those particle species that do \textit{not} have an order parameter. The index $A$ (or $B$, or $C$...) follows Einstein's convention.
Additionally, we need another list of scalars $\varphi^I(x^\alpha)$, which represent the phases of the order parameters of the superfluid species, where also the chemical index $I$ (or $J$, or $K$...) follows Einstein's convention. These fields carry double information. In fact, we can extract the chemical potential from the Josephson relation $\mu^I=-u^\nu \nabla_\nu \varphi^I$, but we can also extract the relativistic analog of the superfluid velocity (as measured in the normal rest frame) from Landau's relation $w^I_\nu =(g\indices{^\rho _\nu}{+}u^\rho u_\nu)\nabla_\rho \varphi^I$ \cite[\S 26]{landau9}. These two relations can be combined to give a unified formula:
\begin{equation}\label{nablphii}
\nabla_\nu \varphi^I =\mu^I u_\nu +w^I_\nu \spc (\text{with } w^I_\nu u^\nu =0) \, .
\end{equation}
Note that, in the non-relativistic limit, $w^I_\nu$ actually becomes $\text{``mass''}\times\text{``superfluid velocity''}$. 

In summary, the hydrodynamic degrees of freedom of a relativistic superfluid are the fields $\Psi(x^\alpha)=\{u^\nu,T,\mu^A,\varphi^I\}$. 

\vspace{-0.2cm}
\subsection{Constitutive relations}\label{costitutzionerobusta}
\vspace{-0.2cm}

The equation of state of the superfluid system is a relation of the form $P=P(T,\mu^A,\mu^I,w^{J\nu}w^I_\nu)$, where $P$ is the thermodynamic pressure. The arguments of $P$ are the available Lorentz scalars that are relevant to thermodynamics. Note that the local values of the phases $\varphi^I$ cannot affect the value of the pressure, as this would break local $U(1)$ symmetry. It was shown in \cite{GusakovAndersson2006} that the differential of $P$ can expressed as follows:
\begin{equation}\label{DPo}
dP=sdT+ n_A d\mu^A +n_I d\mu^I   - \mathcal{H}_{IJ} \,  d\bigg(\dfrac{w^I_\nu w^{J\nu}}{2} \bigg) \, ,
\end{equation}
where $s$ is the entropy density, $n_A$ is the number density of each normal species $A$, $n^I$ the number density of each superfluid species $I$, and $\mathcal{H}_{IJ}=\mathcal{H}_{JI}$ is the relativistic superfluid entrainment matrix, all measured in the normal rest frame. In the case of a single superfluid species, $\mathcal{H}_{IJ}$ reduces to $\text{ ``superfluid mass density''}/\text{``mass''$\,^2$}$. It was also shown in \cite{GusakovAndersson2006} that the entropy current $s^\nu$, particle currents $n_A^\nu$ and $n_I^\nu$ and the stress-energy tensor $T^{\nu \rho}$ take the form
\begin{equation}\label{gusakko}
\begin{split}
s^\nu ={}& s u^\nu \, , \\
n_A^\nu ={}& n_A u^\nu \, , \\
n_I^\nu ={}& n_I u^\nu +\mathcal{H}_{IJ} w^{J\nu} \, , \\
T^{\nu \rho} ={}& (\rho{+}P) u^\nu u^\rho +P g^{\nu \rho} + \mathcal{H}_{IJ} \, \big( \mu^J  u^\nu w^{I\rho} + \mu^I u^\rho w^{J\nu}+ w^{J\nu} w^{I\rho}\big) \, ,\\
\end{split}
\end{equation}
where $\rho =sT+ n_A \mu^A +n_I \mu^I-P$ is the energy density as measured in the normal rest frame (i.e. $\rho=T^{\nu \rho}u_\nu u_\rho$).

\subsection{Equations of motion}\label{eom}
\vspace{-0.2cm}

Let us recall that the fundamental fields of the theory are $\Psi=\{u^\rho,T,\mu^A,\varphi^I\}$. Clearly, each degree of freedom should have a corresponding local equation of motion. The energy-momentum balance law $\nabla_\nu T^{\nu \rho}=f^\rho$ takes care of $T$ and $u^\rho$, where $f^\rho$ is the four-force exerted by the environment on the fluid. Thus, we only need one scalar equation for each chemical component $A$ and $I$. These are just the balance laws $\nabla_\nu n_A^\nu=R_A$ and $\nabla_\nu n_I^\nu=R_I$, where $R_A$ and $R_I$ could be some internal reaction rates or some supply/depletion force coming from the environment. The exact values of $f^\rho$, $R_A$, and $R_I$ may be left unknown for our purpose. However, all should work out together in a way that the total system ``superfluid+environment'' obeys the laws of thermodynamics. For example, we must have that
\begin{equation}\label{thesecondlaw}
\nabla_\nu s^\nu +\nabla_\nu s^\nu_{\text{environment}}\geq 0 \, . 
\end{equation}

\vspace{-0.5cm}
\subsection{Some convenient notation}\label{conveniamo}
\vspace{-0.2cm}

We conclude the section by introducing an elegant notation, due to \citet{Carter_starting_point}, which allows us to rewrite the hydrodynamic equations in a more compact form. First, let us define the following covectors:
\begin{equation}\label{equzione6}
\begin{split}
T_\nu \equiv {}& T u_\nu +\dfrac{1}{s}(\mathcal{H}_{IJ}\mu^J -n_I)w^I_\nu \, , \\
\mu^A_\nu \equiv {}& \mu^A u_\nu \, , \\
\mu^I_\nu \equiv {}& \mu^I u_\nu+w^I_\nu =\nabla_\nu \varphi^I \, . \\
\end{split}
\end{equation}
Then, let us introduce a ``combined'' chemical index
$X=\{s,A,I\}$, which runs over all the four-currents $\{s^\nu,n_A^\nu,n_I^\nu \}$ of the system. With this notation in place, one can show that the differential of the pressure and the constitutive relation of the stress-energy tensor can be recast in a very compact form:
\begin{equation}\label{notazionona}
\begin{split}
dP ={}& -n_X^\nu d\mu^X_\nu \, ,\\
T\indices{^\nu _\rho} ={}& Pg\indices{^\nu _\rho} {+}n_X^\nu \mu^X_\rho \, ,\\
\end{split}
\end{equation}
which will be useful later.
This completes our review of relativistic superfluid hydrodynamics. 

\vspace{-0.3cm}
\section{Stability criterion in general relativity}\label{stabiliycrituzzione}
\vspace{-0.2cm}

As discussed in the introduction, our goal is to identify those local minima of the grand-potential where the superfluid is in motion relative to the environment (see figure \ref{fig:1}). These are flows that do not experience friction with the environment and can potentially last forever. In this section, we work with a generic relativistic superfluid mixture. We identify all those states that make the grand-potential stationary, and we determine necessary and sufficient conditions for these states to be genuine minima. This will result in a set of flow-dependent thermodynamic inequalities, which are analogous to the textbook thermodynamic inequalities on specific heats and compressibilities (see e.g. \cite[\S 21]{landau5} and \cite[\S 12]{prigoginebook}). However, while those standard inequalities determine whether a certain phase of matter is stable, our present inequalities determine whether a flow is long-lived.

\vspace{-0.4cm}
\subsection{Stationary points of the grand-potential}\label{theton}
\vspace{-0.2cm}

In a chemical mixture, the grand potential reads $\Omega=U-T_\star S-\mu_\star^A N_A -
\mu_\star^I N_I$, or equivalently $\Omega =U-\mu_\star^X N_X$ (see section \ref{conveniamo}), where the coefficients $\mu_\star^X=\{T_\star,\mu_\star^A,\mu_\star^I\}$ are background constants, defined as
\begin{equation}
\mu_\star^X =\dfrac{\partial U^\text{environment}}{\partial N_X^{\text{environment}}} \, .
\end{equation}
In the presence of a gravitational field, these parameters coincide with the \textit{redshifted} chemical potentials of the environment \cite[\S 27]{landau5}, which are uniform in spacetime. The extensive quantities $U$ and $N_X$ refer to the superfluid, and they are integrals of the respective currents, where the conserved energy current is built as the contraction $-\mathcal{K}^\rho T\indices{^\nu _\rho}$ \cite[\S 3.8]{carroll_2019}, where $\mathcal{K}^\rho$ is a timelike Killing vector that identifies the local rest frame of the environment at every event. Introducing, for convenience, the fugacities $\alpha_\star^X\,{=}\,\mu_\star^X/T_\star$ and the inverse-temperature four-vector $\beta_\star^\nu\,{=}\,- \mathcal{K}^\nu/T_\star$ \cite{Israel_Stewart_1979,Israel_2009_inbook,BecattiniBeta2016} (which is also a Killing vector), we obtain the following expression for the grand potential:
\begin{equation}
\Omega=-T_\star \int_{\Sigma} \phi^\nu d\Sigma_\nu \spc (\text{orientation: } d\Sigma_0>0)\, ,
\end{equation}
where $\Sigma$ is a spacelike Cauchy surface of interest, and $\phi^\nu {=} \alpha_\star^X n^\nu_X {+}\beta_\star^\rho T\indices{^\nu _\rho} $. We stress again that $\{ \alpha_\star^X, \beta_\star^\nu\}$ are properties \textit{of the environment} (see footnote \ref{foononanan1ona}), and thus they play the role of externally fixed parameters. We use the subscript ``$\star$'' as a reminder that the corresponding quantity does \textit{not} depend on the fluid's state.

\newpage

To find the minima of $\Omega$, we follow the same procedure as in the introduction. We consider a smooth one-parameter family of ``histories'' $\{\Psi (\lambda)\}_{\lambda \in \mathbb{R}}$ at fixed external conditions $\{g^{\alpha \beta},\alpha_\star^X, \beta_\star^\nu \}$. Each value of $\lambda$ corresponds to a different physically admissible realization of the superfluid. We assume that $\lambda=0$ corresponds to a (yet to be determined) local minimum of the grand potential, and we require that $\dot{\Omega}(0)=0$ and $\ddot{\Omega}(0) \geq 0$, where the ``$\dot{\textcolor{white}{m}}$'' denotes differentiation in $\lambda$ (not in time!). The derivatives of $\phi^\nu(\lambda)$ are most easily computed using the notation \eqref{notazionona}:
\begin{equation}\label{phioddottorne}
\begin{split}
\phi^\nu={}&(\alpha_\star^X{+}\beta_\star^\rho \mu^X_\rho) n^\nu_X +P \beta_\star^\nu \, , \\
\dot{\phi}^\nu={}& (\alpha_\star^X{+}\beta_\star^\rho \mu^X_\rho) \dot{n}^\nu_X +2 \beta^{[\rho}_\star n^{\nu]}_X \dot{\mu}^X_\rho \, , \\
\ddot{\phi}^\nu={}& (\alpha_\star^X{+}\beta_\star^\rho \mu^X_\rho) \ddot{n}^\nu_X +2 \beta^{[\rho}_\star n^{\nu]}_X \ddot{\mu}^X_\rho  + (2\beta^{\rho}_\star \dot{n}^{\nu}_X{-}\beta^{\nu}_\star \dot{n}^{\rho}_X) \dot{\mu}^X_\rho \, . \\
\end{split}
\end{equation}
These same formulae were obtained in \cite{GavassinoStabilityCarter2022}. However, in that case, we were looking for the \textit{absolute} minimum of $\Omega$, and we required that $\dot{\Omega}(0)=0$ for completely arbitrary values of $\dot{\mu}^X_\nu(0)$, including states where $\dot{\mu}^I_\nu(\lambda)$ was not a closed differential form, which is forbidden by equation \eqref{equzione6}. This time, to find the \textit{local} minima, we need to be more careful. We must implement the constraint 
$\mu^I_\rho(\lambda) =\nabla_\rho \varphi^I(\lambda)$ explicitly, so that $\dot{\mu}^I_\rho =\nabla_\rho \dot{\varphi}^I$, and $\ddot{\mu}^I_\rho =\nabla_\rho \ddot{\varphi}^I$. This will allow us to find states with non-vanishing superflows. Let us see how this works.

We pick the terms $2 \beta^{[\rho}_\star n^{\nu]}_X \dot{\mu}^X_\rho$ and $2 \beta^{[\rho}_\star n^{\nu]}_X \ddot{\mu}^X_\rho$ in \eqref{phioddottorne}, and we focus on the sum over $I$. Then, we invoke the irrationality constraint to make the following manipulations\footnote{There is a mathematical technicality here. Since the phases $\varphi^I$ are defined modulo $2\pi$, the differential forms $\mu^I_\nu$ are closed, but not exact. Around closed loops like the pipe in figure \ref{fig:1}, one has $\oint \mu^I_\nu dx^\nu =2\pi z$, with $z\in \mathbb{Z}$. Therefore, we are not allowed to make manipulations such as $F\mu^I_\rho=F \nabla_\rho \varphi^I=\nabla_\rho (F\varphi^I)-\varphi^I \nabla_\rho F$, because $\varphi^I$ is multivalued. However, since $\mu^I_\nu (\lambda)$ is continuous in $\lambda$, also the function $z(\lambda)$ is continuous. Combining this with the fact that $z(\lambda)$ can only take integer values, we obtain $0=2\pi \dot{z}=\oint \dot{\mu}^I_\nu dx^\nu$. Thus, the form $\dot{\mu}^I_\nu$ is actually \textit{exact}, meaning that $\dot{\varphi}^I$ (and thus also $\ddot{\varphi}^I$) is a perfectly well-defined single-valued function. Therefore, manipulations such as $F\dot{\mu}^I_\rho =F \nabla_\rho \dot{\varphi}^I=\nabla_\rho (F\dot{\varphi}^I)-\dot{\varphi}^I \nabla_\rho F$ are indeed allowed.}:
\begin{equation}\label{irrez}
\begin{split}
& 2 \beta^{[\rho}_\star n^{\nu]}_I \dot{\mu}^I_\rho =2 \beta^{[\rho}_\star n^{\nu]}_I \nabla_\rho \dot{\varphi}^I =\nabla_\rho (2\dot{\varphi}^I\beta^{[\rho}_\star n^{\nu]}_I)-\dot{\varphi}^I \nabla_{\rho}(\beta^{\rho}_\star n^{\nu}_I{-}\beta^{\nu}_\star n^{\rho}_I)=\nabla_\rho (2\dot{\varphi}^I\beta^{[\rho}_\star n^{\nu]}_I){-}\dot{\varphi}^I[(\mathfrak{L}_{\beta_\star} n_I)^\nu-\beta_\star^\nu \nabla_\rho n_I^\rho ] \, , \\
& 2 \beta^{[\rho}_\star n^{\nu]}_I \ddot{\mu}^I_\rho =2 \beta^{[\rho}_\star n^{\nu]}_I \nabla_\rho \ddot{\varphi}^I =\nabla_\rho (2\ddot{\varphi}^I\beta^{[\rho}_\star n^{\nu]}_I)-\ddot{\varphi}^I \nabla_{\rho}(\beta^{\rho}_\star n^{\nu}_I{-}\beta^{\nu}_\star n^{\rho}_I)=\nabla_\rho (2\ddot{\varphi}^I\beta^{[\rho}_\star n^{\nu]}_I){-}\ddot{\varphi}^I[(\mathfrak{L}_{\beta_\star} n_I)^\nu-\beta_\star^\nu \nabla_\rho n_I^\rho ]\, ,\\
\end{split}
\end{equation}
with $\mathfrak{L}$ the Lie derivative. In the last step, we used the fact that $\nabla_\rho \beta_\star^\rho=g^{\nu \rho}\nabla_{(\nu}\beta_{\star \rho)}=0$. Introducing the short-hand notations $Z_{(1)}^{[\rho \nu]}=2\dot{\varphi}^I\beta^{[\rho}_\star n^{\nu]}_I$ and $Z_{(2)}^{[\rho \nu]}=2\ddot{\varphi}^I\beta^{[\rho}_\star n^{\nu]}_I$, the derivatives of $\phi^\nu(\lambda)$ can be rewritten (for any $\lambda$) as follows:
\begin{equation}\label{thphidot}
\begin{split}
\dot{\phi}^\nu={}& (\alpha_\star^X{+}\beta_\star^\rho \mu^X_\rho) \dot{n}^\nu_X +2 \beta^{[\rho}_\star u^{\nu]}(s\dot{T}_\rho {+}n_A \dot{\mu}^A_\rho) -\dot{\varphi}^I[(\mathfrak{L}_{\beta_\star} n_I)^\nu{-}\beta_\star^\nu \nabla_\rho n_I^\rho ] +\nabla_\rho {Z}_{(1)}^{[\rho \nu]} \, , \\
\ddot{\phi}^\nu={}& (\alpha_\star^X{+}\beta_\star^\rho \mu^X_\rho) \ddot{n}^\nu_X +2 \beta^{[\rho}_\star u^{\nu]}(s\ddot{T}_\rho {+}n_A \ddot{\mu}^A_\rho) -\ddot{\varphi}^I[(\mathfrak{L}_{\beta_\star} n_I)^\nu{-}\beta_\star^\nu \nabla_\rho n_I^\rho ]+ \nabla_\rho {Z}_{(2)}^{[\rho \nu]} + (2\beta^{\rho}_\star \dot{n}^{\nu}_X{-}\beta^{\nu}_\star \dot{n}^{\rho}_X) \dot{\mu}_\rho^X  \, . \\
\end{split}
\end{equation}
When we integrate over the Cauchy surface $\Sigma$, the terms $\nabla_\rho {Z}_{(i)}^{[\rho \nu]}$ become boundary contributions by Gauss' theorem, $\int_\Sigma \! \nabla_\rho Z_{(i)}^{[\rho \nu]} d\Sigma_\nu\,{=}\,\frac{1}{2}\oint_{\partial \Sigma}\! Z_{(i)}^{[\rho \nu]}dS_{\nu \rho}$ \cite[\S 3.3.3]{PoissonToolkit2009pwt}, and thus vanish assuming the densities drop to zero at infinity\footnote{If the superfluid is enclosed in a container with impenetrable walls, then the two-form $Z_{(i)}^{[\rho \nu]}dS_{\nu \rho}$ is automatically zero on the boundary. The proof is the following. Let $\tau_\nu$ be the normal to the Cauchy surface $\Sigma$, and $r_\nu$ the normal to the 3D hypersurface drawn by the wall in spacetime. Then, $dS_{\nu \rho}\propto \tau_\nu r_\rho-\tau_\rho r_\nu  $. But the walls of the container are part of the environment, so they should be at rest relative to it, giving $\beta_\star^\nu r_\nu=0$. Furthermore, impenetrability of the wall means $n_I^\nu r_\nu=0$. Thus, $Z_{(i)}^{[\rho \nu]}r_\nu =0$, since $Z_{(i)}^{[\rho \nu]} \propto \beta^{[\rho}_\star n^{\nu]}_I$.}. Hence, $\dot{\Omega}(0)=0$ only requires that the first three terms in $\dot{\phi}^\nu (0)$ be 0, for all allowed choices of $\{\dot{n}_X^\nu(0),\dot{T}_\rho (0),\dot{\mu}^A_\rho(0),\dot{\varphi}^I(0)\}$. {\color{black}This happens if the corresponding coefficients are zero, i.e. if the following holds (for $\lambda=0$, of course):
\begin{equation}\label{equillo}
\boxed{\beta_\star^\rho \mu^X_\rho=-\alpha_\star^X \, ,} \spc \boxed{\beta^{[\rho}_\star u^{\nu]}=0 \, ,} \spc  \boxed{(\mathfrak{L}_{\beta_\star} n_I)^\nu =\beta_\star^\nu \nabla_\rho n_I^\rho \, . }
\end{equation}
Let us see what these conditions entail.} Since $\alpha_\star^s=T_\star/T_\star=1$, the first two conditions can be equivalently rearranged as $\mu^X/T=\alpha_\star^X=\text{``constant''}$ and $u^\nu/T=\beta_\star^\nu=\text{``Killing vector''}$, which are the usual conditions for the thermal, diffusive, and hydrostatic equilibrium in relativity \cite{Israel1981}. These conditions also imply that $(\mathfrak{L}_{\beta_\star} u)^\nu =\mathfrak{L}_{\beta_\star}\mu^X=0$, which can be verified via direct computation, keeping in mind that $(\mathfrak{L}_{\beta_\star} g)_{\nu \rho}=0$. What is perhaps less obvious is that also the one-forms $d\varphi^I=\mu_\nu^Idx^\nu$ are conserved along the flow generated by $\beta_\star^\nu$. This can be shown using Cartan's magic formula: $\mathfrak{L}_{\beta_\star}d\varphi^I=(d\iota_{\beta_\star}+\iota_{\beta_\star} d)d\varphi^I=d(\beta_\star^\rho \mu^I_\rho)=-d\alpha_\star^I=0$. But then, all physical quantities are invariant along the Killing flow, i.e. $\mathfrak{L}_{\beta_\star} \text{``observables''}=0$, because any hydrodynamic field is built from $\{g_{\nu \rho},u^\nu,T,\mu^A,\mu^I_\nu\}$, all of which are invariant. In particular, we necessarily have that $(\mathfrak{L}_{\beta_\star} n_I)^\nu=0$, so the last equation of \eqref{equillo} becomes $\nabla_\rho n_I^\rho=0$. But we also have that $\nabla_\nu (su^\nu)=\nabla_\nu (n_A u^\nu)=0$ (because $\nabla_\nu u^\nu=\mathfrak{L}_{\beta_\star} n_X=0$). Thus, the flow is reversible, and no net particle transfusion is taking place. Combining everything together, we conclude that, while superfluids do undergo perpetual motion, they cannot be used as ``perpetual machines'', since the long-lived states admit no time dependence in the frame of the environment, and sustain no perpetual chemical or mechanical cycle. 


One can verify that all states fulfilling \eqref{equillo} always solve the equations of motion in section \ref{eom}, with $f^\nu{=}R_I{=}R_A{=}0$. Since $w_\nu^I$ does not need to vanish, the conditions \eqref{equillo} provide a characterization of all those states where a superfluid undergoes perpetual frictionless motion in the rest frame of the bath (of which figure \ref{fig:1} offers an example).

\subsection{Actual minima}\label{theinfonannnnnnnnn}
\vspace{-0.2cm}

The requirement $\dot{\Omega}(0)=0$ allowed us to identify some time-independent hydrodynamic solutions with a superflow. However, we still need to determine whether these are actual local minima of the grand-potential, which requires $\ddot{\Omega}(0)\,{\geq}\, 0$. 
To check this, we evaluate \eqref{thphidot}$_2$ at $\lambda\,{=}\, 0$. Invoking \eqref{equillo}, and integrating out the boundary term, we obtain
\begin{equation}\label{ddphiPhione}
\ddot{\Omega}(0) =-T_\star \int_\Sigma \big[2\beta^{\rho}_\star \,  \dot{n}^{\nu}_X(0){-}\beta^{\nu}_\star \, \dot{n}^{\rho}_X(0) \big] \dot{\mu}_\rho^X (0)d\Sigma_\nu \geq 0 \, .
\end{equation}
We can rewrite this expression in a more standard form. Since $A(\lambda)=A(0)+ \dot{A}(0)\lambda +\frac{1}{2}\ddot{A}(0)\lambda^2+\mathcal{O}(\lambda^3)$, for any physical quantity $A$, we can define the ``first-order variation'' $\delta A:=\dot{A}(0)\lambda$, as is usually done in thermodynamics textbooks. Given that $\delta \Omega =\dot{\Omega}(0)\lambda=0$, we should also introduce the functional $T_\star E:=\frac{1}{2}\ddot{\Omega}(0)\lambda^2$, which is the ``second-order variation'' of $\Omega$. Therefore, if we multiply \eqref{ddphiPhione} by $\lambda^2/2$, we obtain (recall that $u^\nu/T {=}\beta_\star^\nu$ at $\lambda{=}0$)
\begin{equation}\label{againinfona}
\begin{split}
& E=\int_\Sigma E^\nu d\Sigma_\nu \geq 0 \, , \text{ with }\\
& T E^\nu = \delta n^\rho_X \delta \mu^X_\rho \dfrac{u^\nu}{2} -\delta n^\nu_X \delta \mu_\rho^X u^\rho \, .\\
\end{split}
\end{equation}
The vector field $E^\nu$ is the so-called information current \cite{GavassinoCausality2021}. The requirement that $E$ be positive definite for all choices of $\Sigma$ and $\delta \Psi\neq 0$ is equivalent to the requirement that $E^\nu$ be timelike (or lightlike) future-directed for all non-vanishing $\delta \Psi$. Under this condition, there exists a macroscopic grand-potential barrier (as in figure \ref{fig:1}) that prevents the spontaneous formation of hydrodynamic fluctuations that may destroy the superflow. In particular, the following facts are all guaranteed to hold automatically:
\begin{itemize}
\item[(i)] The flow is stable to linear thermodynamic transformations, in a way that is analogous to \cite[\S 21]{landau5} and \cite[\S 12]{prigoginebook}.
\item[(ii)] The flow is covariantly stable to linear hydrodynamic perturbations, in the same sense as \cite{Hishcock1983,GavassinoBounds2023}. Furthermore, the hydrodynamic equations of motion are symmetric-hyperbolic and causal when linearized around the given flow \cite{GavassinoCausality2021,GavassinoSuperluminal2021,GavassinoUniversalityI2023} (note that Theorem 1 of \cite{GavassinoUniversalityI2023} holds also for anisotropic background states).
\item[(iii)] The flow is stable also to linear stochastic fluctuations of hydrodynamic type, in the same sense as \cite{GavassinoFluctuatingBDNK2024vyu}.
\end{itemize}
Hence, \eqref{againinfona} is the stability criterion we were looking for. Unfortunately, \eqref{againinfona} is not enough to guarantee stability of the flow to all possible processes. In fact, the flow is not guaranteed to be stable to fluctuations that are not described by hydrodynamics. For this reason, the criterion \eqref{againinfona} is only a necessary condition for a superflow to be long-lived, which is complementary to more microscopic local criteria, such as the one of Landau. 

We also note that the stability criterion \eqref{againinfona} is formally identical to that provided in \cite{GavassinoStabilityCarter2022}, since the information current looks the same. However, now we are expanding around an equilibrium state with $w^I_\nu \neq 0$. Thus, the thermodynamic inequalities will be much  more complicated, as the equilibrium superflows break local isotropy.

\vspace{-0.2cm}
\subsection{Vector decomposition in the rest frame of the bath}
\vspace{-0.2cm}

If we wish to extract useful thermodynamic inequalities from \eqref{againinfona}, we need to express the latter in the more transparent notation given in section \ref{costitutzionerobusta}. This requires some more manipulations, which are discussed below.

The background velocity $u^\nu$ defines the most natural observer for our purposes. In fact, $u^\nu$ is the local velocity of the thermal bath (since $u^\nu \propto \beta_\star^\nu$), and thus marks the reference frame where usual thermodynamic reasoning takes place. Hence, let us make the decomposition
$TE^\nu =\mathcal{E}_0 u^\nu +\mathfrak{e}^\nu$, with $\mathcal{E}_0=-u_\nu TE^\nu$ and $u_\nu\mathfrak{e}^\nu=0$. The quantities $\mathcal{E}_0$ and $\mathfrak{e}^\nu$ have the dimensions of an energy density and an energy flux. Indeed, they can be interpreted as the local density and flux of \textit{exergy} ($=$ ``usable energy in the bath's frame'' \cite{BecattiniExergy2023}). From \eqref{againinfona}, we find
\begin{equation}\label{eee}
\begin{split}
2\mathcal{E}_0={}&  (g\indices{^\rho _\nu}{+}2u^\rho u_\nu) \delta n_X^\nu \delta \mu^X_\rho \, ,\\
\mathfrak{e}^\nu ={}& (g\indices{^\nu _\lambda} {+} u^\nu u_\lambda)  \delta n_X^\lambda (-u^\rho \delta \mu^X_\rho) \, .\\
\end{split}
\end{equation}
Then, we can rewrite \eqref{eee} by simply invoking \eqref{gusakko} and \eqref{equzione6}. The intermediate manipulations are lengthy, but they are just conventional derivatives, since $\delta A=\dot{A}(0)\lambda$. Here, we only give the final result. Defined the projected vector
$\delta \Bar{w}^J_\nu =(g\indices{^\rho _\nu}{+}u^\rho u_\nu)\delta w^J_\rho$, we have the following expressions:
\begin{equation}\label{ammazzaoh}
\begin{split}
2\mathcal{E}_0={}&  \delta n_X \delta \mu^X +(\rho{+}P)\delta u^\nu \delta u_\nu +\mathcal{H}_{IJ}( \delta \Bar{w}^{I\nu}\delta\Bar{w}^J_\nu {+} w^I_\nu  w^J_\rho \delta u^\nu \delta u^\rho) +\delta \mathcal{H}_{IJ} w^{I\nu}\delta \Bar{w}^J_\nu +2 \delta u^\nu \delta (\mu^I \mathcal{H}_{IJ} w^J_\nu) \, ,\\
\mathfrak{e}^\nu ={}& (n_X \delta \mu^X {+}\mu^I \mathcal{H}_{IJ}w^J_\rho \delta u^\rho)\delta u^\nu +(\delta \mu^I {+}w^I_\rho \delta u^\rho)(\delta \mathcal{H}_{IJ} w^{J\nu}{+}\mathcal{H}_{IJ}\delta \Bar{w}^{J\nu}) \, .\\
\end{split}
\end{equation}

\subsection{Main Theorem}
\vspace{-0.2cm}

Now we are ready to reformulate the stability criterion \eqref{againinfona} in a form that is better suited for applications.

The requirement that $TE^\nu =\mathcal{E}_0 u^\nu +\mathfrak{e}^\nu$ be timelike/lightlike future-directed is equivalent to the statement that any observer $\mathcal{O}$ who measures the zeroth component of $E^\nu$  obtains a non-negative outcome. This, in turn, means that
\begin{equation}\label{rmoeoce}
\mathcal{E}= \dfrac{TE^\nu u_{\mathcal{O}\nu}}{u^\rho u_{\mathcal{O}\rho}} \geq 0 \, ,
\end{equation}
for any four-velocity $u^\nu_\mathcal{O}$. Again, we can perform the ``bath-frame'' decomposition $u_\mathcal{O}^\nu \propto u^\nu {+}V^\nu$, where $V^\nu$ is the ordinary three-velocity of $\mathcal{O}$ relative to the bath, with $u_\nu V^\nu=0$, and $V^\nu V_\nu \in [0,1)$. Then, equation \eqref{rmoeoce} becomes $\mathcal{E}=\mathcal{E}_0-\mathfrak{e}^\nu V_\nu \geq 0$. With the aid of \eqref{ammazzaoh}, we finally obtain the main result of this paper:
{\color{black}
\begin{theorem}\label{theo1}
A superfluid state $\{u^\nu (x^\alpha),T(x^\alpha),\mu^A(x^\alpha),\varphi^I(x^\alpha)\}$ that fulfills \eqref{equillo} is a local minimum of the grand potential $\Omega =U-\mu_\star^X N_X$ only if the quadratic form 
\begin{equation}\label{TheMainE}
\boxed{
\begin{aligned}
2\mathcal{E}={}&  \delta n_X \delta \mu^X +(\rho{+}P)\delta u^\nu \delta u_\nu +\mathcal{H}_{IJ}( \delta \Bar{w}^{I\nu}\delta\Bar{w}^J_\nu {+} w^I_\nu  w^J_\rho \delta u^\nu \delta u^\rho) +\delta \mathcal{H}_{IJ} w^{I\nu}\delta \Bar{w}^J_\nu +2 \delta u^\nu \delta (\mu^I \mathcal{H}_{IJ} w^J_\nu) \\
-{}& 2(n_X \delta \mu^X {+}\mu^I \mathcal{H}_{IJ}w^J_\rho \delta u^\rho)\delta u^\nu V_\nu -2(\delta \mu^I {+}w^I_\rho \delta u^\rho)(\delta \mathcal{H}_{IJ} w^{J\nu} V_\nu{+}\mathcal{H}_{IJ}\delta \Bar{w}^{J\nu} V_\nu) \\
\end{aligned}
}
\end{equation}
is non-negative at every spacetime point, and for all admissible values of $\{\delta \mu^X,\delta u^\nu,\delta w^I_\nu,V^\nu \}$. If $\mathcal{E}$ is strictly positive definite, then statements \textup{(i,ii,iii)} of section \ref{theinfonannnnnnnnn} are fulfilled for linear perturbations around the state of interest.
\end{theorem}}
In Appendix \ref{pinnezzo}, we show that the positivity of \eqref{TheMainE} also guarantees hydrodynamic stability of states that are filled with a lattice of many vortices, provided that the latter are pinned to the environment. If, instead, the vortices are free to move, they become a source of friction, which makes all relative flows unstable \cite{langlois98,GavassinoIordanskii2021}. Then, the only local minimum of $\Omega$ is the full equilibrium state (with $w^I_\nu =0$), and the present analysis reduces to that of \cite{GavassinoStabilityCarter2022}.



\vspace{-0.2cm}
\section{Application to single-component superfluids}\label{RelativisticSingleComponent}
\vspace{-0.2cm}

Let us apply Theorem \ref{theo1} to a simple finite-temperature superfluid, with a single chemical species.

\vspace{-0.3cm}
\subsection{Derivation of the quadratic form}
\vspace{-0.2cm}

For a single-component superfluid, the ``chemical index'' $X$ runs over two values: $X\in \{s,n\}$, where $s$ is the entropy and $n$ is the particle number. There is no chemical potential of ``$A$-type''. The species $n$ has an associated order parameter $\varphi$, with gradient $\nabla_\nu \varphi=\mu u_\nu +w_\nu $ . Hence, according to \eqref{DPo}, the differential of the pressure reads
\begin{equation}\label{dPippo}
dP= sdT + n d\mu   - \mathcal{H} \,  d\bigg(\dfrac{w_\nu w^{\nu}}{2} \bigg) \, ,
\end{equation}
and the exergy function \eqref{TheMainE} becomes
\begin{equation}\label{jkmwkoodwm}
\begin{split}
2\mathcal{E}={}& \delta s \delta T + \delta n \delta \mu +h \, \delta u^\nu \delta u_\nu +\mathcal{H}( \delta \Bar{w}^{\nu}\delta\Bar{w}_\nu {+} w_\nu  w_\rho \delta u^\nu \delta u^\rho) +\delta \mathcal{H} w^{\nu}\delta \Bar{w}_\nu +2 \delta u^\nu \delta (\mu \mathcal{H} w_\nu) \\
-{}& 2(s\delta T +n \delta \mu {+}\mu \mathcal{H}w_\rho \delta u^\rho)\delta u^\nu V_\nu -2(\delta \mu {+}w_\rho \delta u^\rho)(\delta \mathcal{H} w^{\nu} V_\nu{+}\mathcal{H}\delta \Bar{w}^{\nu} V_\nu) \, , \\
\end{split}    
\end{equation}
where we introduced the short-hand notation $h =\rho{+}P$.
To extract inequalities from \eqref{jkmwkoodwm}, we can interpret $\mathcal{E}$ as a quadratic form in the variables $\{\delta T, \delta \mu, \delta u^\nu, \delta \Bar{w}_\nu \}$. This requires that we decompose the variations of all other variables as linear combinations of these primary variables, namely
\begin{equation}\label{diffenrzioni}
\begin{split}
\delta s ={}& s_T \delta T + n_T \delta \mu -\mathcal{H}_T \delta\bigg(\dfrac{w_\nu w^{\nu}}{2} \bigg) \, ,\\
\delta n ={}& n_T \delta T + n_\mu \delta \mu -\mathcal{H}_\mu \delta \bigg(\dfrac{w_\nu w^{\nu}}{2} \bigg) \, ,\\
\delta \mathcal{H} ={}& \mathcal{H}_T \delta T + \mathcal{H}_\mu \delta \mu +\mathcal{H}_w \delta \bigg(\dfrac{w_\nu w^{\nu}}{2} \bigg)\, ,\\
\end{split}    
\end{equation}
with $\delta(w_\nu w^\nu/2)=w^\nu \delta \Bar{w}_\nu$. To derive \eqref{diffenrzioni}, we used all the Maxwell relations that \eqref{dPippo} entails, specifically $s_\mu =n_T$, $s_w =-\mathcal{H}_T$, $n_w=-\mathcal{H}_\mu$. Thus, \eqref{jkmwkoodwm} becomes
\begin{equation}
\begin{split}
2\mathcal{E}={}& s_T (\delta T)^2 +2 n_T \delta T \delta \mu +n_\mu (\delta \mu)^2 
+ h\,\delta u^\nu \delta u_\nu + \mathcal{H}( \delta \Bar{w}^{\nu}\delta\Bar{w}_\nu {+} w_\nu  w_\rho \delta u^\nu \delta u^\rho) +\mathcal{H}_w  (w^{\nu}\delta \Bar{w}_\nu)^2 \\ 
+{}& 2 \delta u^\nu  \big[\delta \mu \mathcal{H} w_\nu+\mu \mathcal{H} \delta \Bar{w}_\nu + \mu \mathcal{H}_T \delta T  w_\nu + \mu \mathcal{H}_\mu \delta \mu  w_\nu +\mu \mathcal{H}_w  w^\rho \delta \Bar{w}_\rho w_\nu \big] 
- 2(s\delta T +n \delta \mu {+}\mu \mathcal{H}w_\rho \delta u^\rho)\delta u^\nu V_\nu \\ -{}& 2(\delta \mu {+}w_\rho \delta u^\rho)(\mathcal{H}_T \delta T w^{\nu} V_\nu + \mathcal{H}_\mu \delta \mu w^{\nu} V_\nu +\mathcal{H}_w  w^\lambda \delta \Bar{w}_\lambda w^{\nu} V_\nu{+}\mathcal{H}\delta \Bar{w}^{\nu} V_\nu) \, . \\
\end{split}    
\end{equation}

\subsection{Matrix representation of the quadratic form}\label{laquadratazza}
\vspace{-0.2cm}

Fix an event $\mathscr{P}$, and work in a local Lorentz frame such that $u^\nu{=}(1,0,0,0)$, 
 $w_\nu{=}(0,w,0,0)$, and $V_\nu {=}-(0,V,W,0)$. Then, we have that $\delta u^\nu =(0,\delta u^1,\delta u^2, \delta u^3)$ and $\delta \Bar{w}_\nu=(0,\delta w_1, \delta w_2, \delta w_3)$. Hence, the independent degrees of freedom of a perturbation at a point are 
$\Psi=\{\delta T, \delta \mu , \delta u^1, \delta w^1 ,\delta u^2 , \delta w^2, \delta u^3, \delta w^3\}$, and there is no algebraic constraint on them. This allows us to write $2\mathcal{E}=\Psi^T (\mathbb{M}_0 + \mathbb{M}_1 V + \mathbb{M}_2 W)\Psi$, where $\mathbb{M}_i$ are three $8\times 8$ symmetric matrices, given by
\begin{equation}
\mathbb{M}_0 =
\left[
\begin{array}{@{}cccc|cc|cc@{}}
s_T & s_\mu & \mu \mathcal{H}_T w & 0 & 0 & 0 & 0 & 0\\
n_T & n_\mu & (\mathcal{H}{+}\mu \mathcal{H}_\mu)w & 0 & 0 & 0 & 0 & 0\\
\mu \mathcal{H}_T w & (\mathcal{H}{+}\mu \mathcal{H}_\mu)w & h{+}\mathcal{H}w^2 & \mu (\mathcal{H}{+}\mathcal{H}_w w^2) & 0 & 0 & 0 & 0\\
0 & 0 & \mu (\mathcal{H}{+}\mathcal{H}_w w^2) & \mathcal{H}{+}\mathcal{H}_w w^2 & 0 & 0 & 0 & 0\\
\hline
0 & 0 & 0 & 0 & h & \mu \mathcal{H} & 0 & 0\\
0 & 0 & 0 & 0 & \mu \mathcal{H} & \mathcal{H} & 0 & 0\\
\hline
0 & 0 & 0 & 0 & 0 & 0 & h & \mu \mathcal{H}\\
0 & 0 & 0 & 0 & 0 & 0 & \mu \mathcal{H} & \mathcal{H} \\
\end{array}
\right] \, ,
\end{equation}
\begin{equation}
\mathbb{M}_1 =
\left[
\begin{array}{@{}cccc|cc|cc@{}}
0 & \mathcal{H}_T w & s {+}\mathcal{H}_T w^2 & 0 & 0 & 0 & 0 & 0\\
\mathcal{H}_T w & 2\mathcal{H}_\mu w & n{+}\mathcal{H}_\mu w^2 & \mathcal{H}{+} \mathcal{H}_w  w^2 & 0 & 0 & 0 & 0\\
s{+}\mathcal{H}_T w^2 & n{+}\mathcal{H}_\mu w^2 & 2\mu \mathcal{H}w & (\mathcal{H}{+} \mathcal{H}_w  w^2)w & 0 & 0 & 0 & 0\\
0 & \mathcal{H}{+} \mathcal{H}_w  w^2 & (\mathcal{H}{+} \mathcal{H}_w  w^2)w & 0 & 0 & 0 & 0 & 0\\
\hline
0 & 0 & 0 & 0 & 0 & 0 & 0 & 0\\
0 & 0 & 0 & 0 & 0 & 0 & 0 & 0\\
\hline
0 & 0 & 0 & 0 & 0 & 0 & 0 & 0\\
0 & 0 & 0 & 0 & 0 & 0 & 0 & 0\\
\end{array} 
\right] \, ,
\end{equation}
\begin{equation}
\mathbb{M}_2 =
\left[
\begin{array}{@{}cccc|cc|cc@{}}
0 & 0 & 0 & 0 & s & 0 & 0 & 0\\
0 & 0 & 0 & 0 & n & \mathcal{H} & 0 & 0\\
0 & 0 & 0 & 0 & \mu \mathcal{H}w & \mathcal{H}w & 0 & 0\\
0 & 0 & 0 & 0 & 0 & 0 & 0 & 0\\
\hline
s & n & \mu \mathcal{H}w & 0 & 0 & 0 & 0 & 0\\
0 & \mathcal{H} & \mathcal{H}w & 0 & 0 & 0 & 0 & 0\\
\hline
0 & 0 & 0 & 0 & 0 & 0 & 0 & 0\\
0 & 0 & 0 & 0 & 0 & 0 & 0 & 0\\
\end{array} 
\right] \, .
\end{equation}
According to Theorem \ref{theo1}, the given state is thermodynamically and hydrodynamically stable at $\mathscr{P}$ if and only if the composite matrices $\mathbb{M}_0{+}\mathbb{M}_1 V {+}\mathbb{M}_2 W$ are positive definite for all choices of $\{ V,W\}$ such that $V^2 {+}W^2 < 1$. The necessary and sufficient conditions for this to happen are rather lengthy to write explicitly, and not so illuminating from a physics standpoint. For practical applications, one should just pick an equation of state of interest, write the matrix $\mathbb{M}_0{+}\mathbb{M}_1 V {+}\mathbb{M}_2 W$ explicitly, and check (numerically) whether it is positive definite.

Here, we just point out some immediately evident facts that follow from the positivity of $\mathbb{M}_0$, which corresponds to the case with $V=W=0$. From the last $2\times 2$ block, we get $h>\mu^2 \mathcal{H}>0$, which is just the statement that the kinetic energy of the superfluid system should be non-negative. From the first $2\times 2$ block, we get
\begin{equation}
\begin{split}
& \dfrac{\partial n}{\partial \mu}\bigg|_{T,w}>0\, , \\
& \dfrac{\partial (s,n)}{\partial (T,\mu)}\bigg|_{w}= \dfrac{\partial (s,n)}{\partial (T,n)}\bigg|_{w}\dfrac{\partial (T,n)}{\partial (T,\mu)}\bigg|_{w}=\dfrac{\partial s}{\partial T}\bigg|_{n,w} \dfrac{\partial n}{\partial \mu}\bigg|_{T,w} >0\, .  \\
\end{split}
\end{equation}
These thermodynamic inequalities are identical to the corresponding ones in normal substances \cite{landau5}, with the additional caveat that one needs to hold $w$ constant in the differentiation. A couple of novel inequalities that are specific to states with a non-vanishing superflow are
\begin{equation}
\begin{split}
& \dfrac{\partial (\mathcal{H}w)}{\partial w}\bigg|_{T,\mu} >0 \, , \\
& \dfrac{\partial s}{\partial T}\bigg|_{\mu, w} > \dfrac{\mu^2 w^2}{h{+}\mathcal{H}w^2} \bigg( \dfrac{\partial \mathcal{H}}{\partial T}\bigg|_{\mu,w} \bigg)^2 \, . \\
\end{split}
\end{equation}

\section{Non-relativistic case}\label{nonrellimitsect}
\vspace{-0.2cm}

There are two possible ways of identifying the local minima of $\Omega$ in a non-relativistic setting. One is to repeat the analysis of section \ref{stabiliycrituzzione} for the theory of \citet{AndreevBash1975}. The other way is to take the results of section \ref{stabiliycrituzzione}, and compute their non-relativistic limit. Here, we choose the latter.

\vspace{-0.3cm}
\subsection{Non-relativistic limit: the procedure}
\vspace{-0.2cm}

To take the non-relativistic limit of our analysis, one needs to make the following approximations \cite{rezzolla_book,GusakovAndersson2006}:
\begin{equation}
\begin{split}
u^\nu \approx {}& 
\begin{pmatrix}
1{+}\frac{1}{2}u^ku_k {-} \phi \\
u^j \\
\end{pmatrix} \, , \\
\mu^X \approx{}& m^X (1+\Bar{\mu}^X) \spc \spc (\text{for }X\neq s)\, , \\
w^I_\nu \approx{}& m^I V^I_\nu \, , \\
\mathcal{H}_{IJ} \approx{}& \dfrac{\rho_{IJ}}{m^I m^J} \, , \\
\rho{+}P \approx{}& \rho \, , \\
\end{split}
\end{equation}
where $\phi$ is the Newtonian gravitational potential, $\Bar{\mu}^X$ are the non-relativistic chemical potentials per unit mass, $m^X$ are the rest masses, $V^I_\nu$ is the superfluid velocity in the normal rest frame, and $\rho_{IJ}$ is the matrix of superfluid mass densities. Additionally, it is convenient to define the rest mass densities $\rho^X{=}m^Xn_X$ (for $X {\neq} s$), which add up to $\rho$. Note that we are still working in units where $c=1$.

\vspace{-0.3cm}
\subsection{Stability criterion}
\vspace{-0.2cm}

The non-relativistic limit of \eqref{equillo}$_1$ and \eqref{equillo}$_2$ is well-known, and is discussed, e.g., in \citet[\S 27]{landau5}.
The non-relativistic limit of \eqref{equillo}$_3$ is straightforward, and does not need additional comments. Instead, the non-relativistic limit of \eqref{TheMainE} deserves more attention. In fact, many terms disappear (the terms $\propto V^\nu$), which are intrinsically relativistic. At the end of the day, we have the following result.{\color{black}
\begin{theorem}\label{Theo2} In the non-relativistic limit, a superfluid state $\{u^k (x^\alpha),T(x^\alpha),\Bar{\mu}^A(x^\alpha),\varphi^I(x^\alpha)\}$ that fulfills the Newtonian analog of \eqref{equillo} is a local minimum of the grand potential $\Omega=U-\mu^X_\star N_X$ only if, in the local rest frame of the environment, the quadratic form 
\begin{equation}\label{TheMainENewt}
\boxed{2\mathcal{E}= \delta s \delta T+ \delta \rho_A \delta \Bar{\mu}^A +\delta \rho_I \delta \Bar{\mu}^I+ \rho\,\delta u^k \delta u_k +\rho_{IJ} \delta V^{Ik}\delta V^J_k  +\delta \rho_{IJ} V^{Ik}\delta V^J_k +2 \delta u^k \sum\nolimits_I  \delta ( \rho_{IJ} V^J_k)  }
\end{equation}
is non-negative at every spacetime point, and for all values of $\{\delta T,\delta \Bar{\mu}^A,\delta \Bar{\mu}^I,\delta u^k,\delta V^I_k \}$. If $\mathcal{E}$ is strictly positive definite, then statements \textup{(i,iii)} of section \ref{theinfonannnnnnnnn} are fulfilled in a neighborhood of the state of interest.
\end{theorem}}
Note that statement (ii) no longer holds in the non-relativistic limit, because we lose information about causality. Also, note that \eqref{TheMainENewt} is calculated in the local rest frame of the environment by construction. 

\vspace{-0.3cm}
\subsection{Single-component superfluid}
\vspace{-0.2cm}

Analogously to what we did in section \ref{RelativisticSingleComponent}, let us specialize Theorem \ref{Theo2} to a simple superfluid, with only one chemical component. In this case, there is no chemical potential with index of type $A$, and a single order-parameter phase $\varphi$. Following standard notation, we call $V_S^k (\equiv V^{Ik})$ the corresponding superfluid velocity in the normal frame, and $\rho_S(\equiv \rho_{IJ})$ the superfluid mass density. Then, equation \eqref{TheMainENewt} becomes
\begin{equation}\label{TheMainENewt33}
2\mathcal{E}= \delta s \delta T +\delta \rho \delta \Bar{\mu}+ \rho\,\delta u^k \delta u_k +\rho_S \delta V^{k}_S\delta V_{Sk}  +\delta \rho_S V^{k}_S\delta V_{Sk} +2 \delta u^k   \delta ( \rho_S V_{Sk})  \, .
\end{equation}
In appendix \ref{comparisonwithandreev}, we show that the stability conditions that follow from \eqref{TheMainENewt33} must coincide with those provided in \cite{Andreev2004}, with the conceptual difference that we have now derived them perturbing around an arbitrary flow fulfilling \eqref{equillo}.

For practical purposes, it is convenient to interpret $\mathcal{E}$ as a quadratic form in the primary variables $\{\delta T,\delta \rho, \delta u^k,\delta V_S^k \}$. Hence, the natural thermodynamic potential of the problem is not the pressure, but the Helmholtz free energy $\mathcal{F}=\rho \Bar{\mu}-P$, whose differential reads
\begin{equation}\label{dPippoNewt}
d\mathcal{F}= -s\, dT + \Bar{\mu} d\rho  + \rho_S \,  d\bigg(\dfrac{V_{Sk} V_S^{k}}{2} \bigg) \, .
\end{equation}
This equation (and the Maxwell relations that follow from it) allows us to write the following decompositions:
\begin{equation}
\begin{split}
\delta s ={}& \dfrac{c_V}{T} \delta T -\partial_T \Bar{\mu}\,  \delta \rho -\partial_T \rho_S \, \delta \bigg(\dfrac{V_{Sk} V_S^{k}}{2} \bigg)  \, , \\
\delta \Bar{\mu} ={}& \partial_T \Bar{\mu} \,\delta T +\partial_\rho \Bar{\mu} \, \delta \rho +\partial_\rho \rho_S \, \delta \bigg(\dfrac{V_{Sk} V_S^{k}}{2} \bigg)  \, , \\
\delta \rho_S ={}& \partial_T \rho_S \, \delta T +\partial_\rho \rho_S \, \delta \rho +\dfrac{\partial_{V_S}\rho_S}{V_S} \delta \bigg(\dfrac{V_{Sk} V_S^{k}}{2} \bigg) \, . \\
\end{split}
\end{equation}
For convenience, we have introduced the specific heat $c_V$ at constant volume \textit{and} superfluid velocity. Furthermore, we have called ``$V_S$'' the norm of the vector $V_S^k$. Then, equation \eqref{TheMainENewt33} becomes
\begin{equation}\label{TheMainENewt44}
\begin{split}
2\mathcal{E}={}& \dfrac{c_V}{T} (\delta T)^2+ \partial_\rho \Bar{\mu} \, (\delta \rho)^2 +\rho\,\delta u^k \delta u_k+2  \rho_S \delta u^k   \delta  V_{Sk} +\rho_S \delta V^{k}_S\delta V_{Sk}   +2\partial_\rho \rho_S \, \delta \rho \, V_S^k \delta V_{Sk} \\
+{}& \dfrac{\partial_{V_S}\rho_S}{V_S}  (V^{k}_S\delta V_{Sk})^2 +2 V_{Sk} \delta u^k   \bigg( \partial_T \rho_S \, \delta T +\partial_\rho \rho_S \, \delta \rho +\dfrac{\partial_{V_S}\rho_S}{V_S} V_S^k \delta V_{Sk} \bigg)  \, .\\
\end{split}
\end{equation}
As we did in section \ref{laquadratazza}, let us orient the space axes such that $V_S^k=(V_S,0,0)$. Then, equation \eqref{TheMainENewt44} takes the form $2\mathcal{E}=\Psi^T \mathbb{M}\Psi$, with $\Psi=\{\delta T, \delta\rho, \delta u^1, \delta V_S^1, \delta u^2, \delta V_S^2,\delta u^3, \delta V_S^3\}$, and
\begin{equation}
\mathbb{M} =
\left[
\begin{array}{@{}cccc|cc|cc@{}}
c_V/T & 0 & \partial_T J_S & 0 & 0 & 0 & 0 & 0\\
0 & \partial_\rho \Bar{\mu} & \partial_\rho J_S &  \partial_\rho J_S & 0 & 0 & 0 & 0\\
\partial_T J_S & \partial_\rho J_S & \rho & \partial_{V_S}J_S & 0 & 0 & 0 & 0\\
0 &  \partial_\rho J_S & \partial_{V_S}J_S &  \partial_{V_S}J_S & 0 & 0 & 0 & 0\\
\hline
0 & 0 & 0 & 0 & \rho & \rho_S & 0 & 0\\
0 & 0 & 0 & 0 & \rho_S & \rho_S & 0 & 0\\
\hline
0 & 0 & 0 & 0 & 0 & 0 & \rho & \rho_S \\
0 & 0 & 0 & 0 & 0 & 0 & \rho_S & \rho_S \\
\end{array} 
\right] \, ,
\end{equation}
where $J_S=\rho_S V_S$ is the mass flux in the normal rest frame. For the flow to be long-lived, this matrix should be positive definite. This produces a list of thermodynamic inequalities. For example, the positivity of the two isolated $2 \times 2$ blocks produces the conditions $\rho>\rho_S>0$, which means that both the superfluid density $\rho_S$ and the normal density $\rho-\rho_S$ must be positive. For the remaining $4\times 4$ block to be positive definite, all the diagonal elements need to be non-negative, which gives the inequalities $c_V>0$, $\partial_\rho \Bar{\mu}>0$, and $ \partial_{V_S}J_S>0$. Additionally, we have the following determinant inequalities: 
\begin{equation}
\label{eq:StabilitySingle}
\begin{split}
& \rho > \bigg(\dfrac{\partial J_S}{\partial V_S}\bigg)_{T,\rho} >0 \, , \\
& \bigg( \dfrac{\partial \Bar{\mu}}{\partial\rho}\bigg)_{T,V_S} \bigg(\dfrac{\partial J_S}{\partial V_S} \bigg)_{T,\rho} > \bigg(\dfrac{\partial J_S}{\partial \rho} \bigg)^2_{T,V_S} \, ,\\
& \dfrac{c_V}{T}\bigg[ \rho-\bigg(\dfrac{\partial J_S}{\partial V_S} \bigg)_{T,\rho} \bigg] > \bigg(\dfrac{\partial J_S}{\partial T} \bigg)^2_{\rho,V_S} \, . \\
\end{split}
\end{equation}
If the superflow is thermodynamically stable, all these inequalities must be fulfilled, which are the same as in \cite{Andreev2004}.

\subsection{Neutron-star models}

Let us now specialize Theorem \ref{Theo2} to a superfluid neutron star, comprised of normal protons and superfluid neutrons. In this case, the thermodynamic state of matter is characterized by a proton chemical potential $\Bar{\mu}^p$ of ``type $A$'' and a neutron chemical potential $\Bar{\mu}^n$ of ``type $I$''. Again, we call $V_S^k (\equiv V^{Ik})$ the superfluid velocity of the neutron component in the normal frame, and $\rho_S(\equiv \rho_{IJ})$ the superfluid mass density of the neutrons. Then, the quadratic form \eqref{TheMainENewt} becomes
\begin{equation}\label{NeutronStar}
2\mathcal{E}= \delta s \delta T+ \delta \rho_p \delta \Bar{\mu}^p +\delta \rho_n \delta \Bar{\mu}^n+ \rho\,\delta u^k \delta u_k +\rho_{S} \delta V^{k}_S\delta V_{Sk}  +\delta \rho_{S} V^{k}_S\delta V_{Sk} +2 \delta u^k  \delta ( \rho_{S} V_{Sk})  \, .
\end{equation}
Again, the most natural primary variables are $\{ \delta T, \delta \rho_p,\delta \rho_n,\delta u^k,\delta V_S^k\}$, and the most natural thermodynamic potential to work with is the free energy, whose differential reads
\begin{equation}\label{dPlutarco}
d\mathcal{F}= -s\, dT + \Bar{\mu}^p d\rho_p+ \Bar{\mu}^n d\rho_n  + \rho_S \,  d\bigg(\dfrac{V_{Sk} V_S^{k}}{2} \bigg) \, .
\end{equation}
Using the Maxwell relations that arise from the above differential, we can write
\begin{equation}
\begin{split}
\delta s ={}& \dfrac{c_V}{T} \delta T -\partial_T \Bar{\mu}^p \,  \delta \rho_p-\partial_T \Bar{\mu}^n \,  \delta \rho_n -\partial_T \rho_S \, \delta \bigg(\dfrac{V_{Sk} V_S^{k}}{2} \bigg)  \, , \\
\delta \Bar{\mu}^p ={}& \partial_T \Bar{\mu}^p \,\delta T +\partial_{\rho_p} \Bar{\mu}^p \, \delta \rho_p+\partial_{\rho_p} \Bar{\mu}^n \, \delta \rho_n +\partial_{\rho_p} \rho_S \, \delta \bigg(\dfrac{V_{Sk} V_S^{k}}{2} \bigg)  \, , \\
\delta \Bar{\mu}^n ={}& \partial_T \Bar{\mu}^n \,\delta T +\partial_{\rho_p} \Bar{\mu}^n \, \delta \rho_p+\partial_{\rho_n} \Bar{\mu}^n \, \delta \rho_n +\partial_{\rho_n} \rho_S \, \delta \bigg(\dfrac{V_{Sk} V_S^{k}}{2} \bigg)  \, , \\
\delta \rho_S ={}& \partial_T \rho_S \, \delta T +\partial_{\rho_p} \rho_S \, \delta \rho_p +\partial_{\rho_n} \rho_S \, \delta \rho_n +\dfrac{\partial_{V_S}\rho_S}{V_S} \delta \bigg(\dfrac{V_{Sk} V_S^{k}}{2} \bigg) \, , \\
\end{split}
\end{equation}
and equation \eqref{NeutronStar} becomes
\begin{equation}\label{NeutronStar2}
\begin{split}
2\mathcal{E}={}& \dfrac{c_V}{T} (\delta T)^2  +\partial_{\rho_p} \Bar{\mu}^p \, (\delta \rho_p)^2+2\partial_{\rho_p} \Bar{\mu}^n \, \delta \rho_n \delta \rho_p+\partial_{\rho_n} \Bar{\mu}^n \, (\delta \rho_n)^2 \\ 
&+ \rho\,\delta u^k \delta u_k+2 \rho_{S} \delta u^k  \delta  V_{Sk}+ \rho_{S} \delta V^{k}_S\delta V_{Sk}+2(\partial_{\rho_p} \rho_S  \delta \rho_p +\partial_{\rho_n} \rho_S  \delta \rho_n) V_{Sk} \delta  V_S^{k}  \\ 
& +\dfrac{\partial_{V_S}\rho_S}{V_S}(V_{Sk} 
 \delta V_S^{k})^2+2 V_{Sk} \delta u^k  \bigg( \partial_T \rho_S \, \delta T +\partial_{\rho_p} \rho_S \, \delta \rho_p +\partial_{\rho_n} \rho_S \, \delta \rho_n +\dfrac{\partial_{V_S}\rho_S}{V_S}V_{Sk} 
 \delta V_S^{k} \bigg) \, . \\
\end{split}  
\end{equation}
As we did in the single-component case, let us work in a reference frame such that $V_S^k=(V_S,0,0)$. Then, we can write $2\mathcal{E}=\Psi^T \mathbb{M}\Psi$, with $\Psi=\{\delta T, \delta\rho_p,\delta \rho_n, \delta u^1, \delta V_S^1, \delta u^2, \delta V_S^2,\delta u^3, \delta V_S^3\}$, and
\begin{equation}
\mathbb{M} =
\left[
\begin{array}{@{}ccccc|cc|cc@{}}
c_V/T & 0 & 0 & \partial_T J_S & 0 & 0 & 0 & 0 & 0\\
0 & \partial_{\rho_p} \Bar{\mu}^p & \partial_{\rho_p} \Bar{\mu}^n & \partial_{\rho_p} J_S & \partial_{\rho_p} J_S & 0 & 0 & 0 & 0 \\
0 & \partial_{\rho_p} \Bar{\mu}^n & \partial_{\rho_n} \Bar{\mu}^n & \partial_{\rho_n} J_S &  \partial_{\rho_n} J_S & 0 & 0 & 0 & 0\\
\partial_T J_S & \partial_{\rho_p} J_S& \partial_{\rho_n} J_S & \rho & \partial_{V_S}J_S & 0 & 0 & 0 & 0\\
0 & \partial_{\rho_p} J_S & \partial_{\rho_n} J_S & \partial_{V_S}J_S &  \partial_{V_S}J_S & 0 & 0 & 0 & 0\\
\hline
0 & 0& 0 & 0 & 0 & \rho & \rho_S & 0 & 0\\
0 & 0& 0 & 0 & 0 & \rho_S & \rho_S & 0 & 0\\
\hline
0 & 0& 0 & 0 & 0 & 0 & 0 & \rho & \rho_S \\
0 & 0&  0 & 0 & 0 & 0 & 0 & \rho_S & \rho_S \\
\end{array} 
\right] \, .
\end{equation}
This matrix is similar to the matrix associated with a single-component superfluid, with the only difference being that there are now two rows and columns associated with the mass densities, instead of just one. This implies that most of the inequalities we obtained in the previous subsection are essentially unchanged (up to some label $p$ or $n$):
\vspace{-0.1cm}
\begin{equation}\label{neutronstore}
\begin{split}
& \rho > \rho_S >0 \, , \\
& \rho > \bigg(\dfrac{\partial J_S}{\partial V_S}\bigg)_{T,\rho_p,\rho_n} >0 \, , \\
& \bigg( \dfrac{\partial \Bar{\mu}^n}{\partial\rho_n}\bigg)_{T,\rho_p,V_S} \bigg(\dfrac{\partial J_S}{\partial V_S} \bigg)_{T,\rho_p,\rho_n} > \bigg(\dfrac{\partial J_S}{\partial \rho_n} \bigg)^2_{T,\rho_p,V_S} \, ,\\
& \bigg( \dfrac{\partial \Bar{\mu}^p}{\partial\rho_p}\bigg)_{T,\rho_n,V_S} \bigg(\dfrac{\partial J_S}{\partial V_S} \bigg)_{T,\rho_p,\rho_n} > \bigg(\dfrac{\partial J_S}{\partial \rho_p} \bigg)^2_{T,\rho_n,V_S} \, ,\\
& \dfrac{c_V}{T}\bigg[ \rho-\bigg(\dfrac{\partial J_S}{\partial V_S} \bigg)_{T,\rho_p,\rho_n} \bigg] > \bigg(\dfrac{\partial J_S}{\partial T} \bigg)^2_{\rho_p,\rho_n,V_S} \, . \\
\end{split}
\end{equation}
There is however one additional inequality, which is the result of having two density variables that may fluctuate in different ways (all partial derivatives are understood to be in the variables $T,\rho_p,\rho_n,V_S$):
\begin{equation}\label{neutronstore2}
\big[\partial_{\rho_p} \Bar{\mu}^p \, \partial_{\rho_n} \Bar{\mu}^n-(\partial_{\rho_p} \Bar{\mu}^n)^2\big]\partial_{V_S} J_S > \partial_{\rho_p} \Bar{\mu}^p (\partial_{\rho_n} J_S)^2+2 \partial_{\rho_p} \Bar{\mu}^n \, \partial_{\rho_p} J_S \, \partial_{\rho_n} J_S+\partial_{\rho_n} \Bar{\mu}^n (\partial_{\rho_p} J_S)^2\, .
\end{equation}
Quantitative applications of these inequalities to superfluid neutron stars will be discussed in future articles.

\section{Conclusions}

By applying fundamental thermodynamic principles, we have derived necessary conditions for a hydrodynamic state with a non-vanishing superflow to be long-lived and stable against all forms of thermodynamic fluctuations, including spontaneous heat exchange, acceleration, and chemical imbalance. Specifically, our analysis provides a universal geometric characterization of all the flows that correspond to stationary points of the grand potential $\Omega=U{-}T_\star S{-}\mu_\star^A N_A {-}
\mu_\star^I N_I$ (see equation \eqref{equillo}), as well as the thermodynamic criteria that ensure such stationary points are true minima of $\Omega$, both in the relativistic case (Theorem \ref{theo1}) and in the Newtonian limit (Theorem \ref{Theo2}). 

{\color{black}The stationary-point conditions \eqref{equillo} align with physical expectations and previous studies \cite{khalatnikov_book,Termo}. They establish that a flow geometry can be long-lived and frictionless \textit{only if} it fulfills the conditions below:
\begin{itemize}
\item The normal component is at rest relative to the environment, exhibiting neither shear nor expansion;
\item The redshifted temperature and chemical potentials remain uniform across the whole body of the superfluid, and they coincide with those of the environment;
\item The flow is time-independent in the rest frame of the environment, with no net chemical exchange or entropy production.
\end{itemize}
The above criteria hold regardless of the shape of the container, and remain valid even in the presence of a vortex array, as long as all vortices are pinned to the environment.

Requiring that a given stationary point of $\Omega$ correspond to a true minimum (as opposed to a maximum, or saddle) was found to be equivalent to demanding that the quadratic form \eqref{TheMainE} [or \eqref{TheMainENewt}, in the Newtonian limit] be positive definite everywhere along the flow. This quadratic form naturally decomposes into three contributions, each carrying a distinct physical meaning:
\begin{itemize}
\item The ``thermodynamic term'' (i.e. the piece $\delta n_X \delta \mu^X$) ensures stability against local temperature and density fluctuations. Its positivity leads to the standard ``textbook'' inequalities on specific heats and compressibilities that are required for the stability of any phase of matter \cite[\S 21]{landau5}.
\item  The ``kinetic term'' (i.e. all pieces with $\delta u\delta u$, $\delta u \delta w^I$, $\delta w^I\delta w^J$) ensures stability against spontaneous accelerations/decelerations of the normal or superfluid components\footnote{\color{black}We know that a superfluid flowing in a circular pipe (as in figure \ref{fig:1}) cannot undergo a uniform slowdown, since the circulation integral $\oint \mu_\theta^I d\theta$ is conserved, implying that $\oint \delta \mu_\theta^I d\theta=0$. Nevertheless, the flow may still undergo localized decelerations compensated by accelerations localised elsewhere (for instance $\delta\mu^I_\theta=A \sin(\theta)$). For the unperturbed flow to be long-lived, such fluctuations must not reduce $\Omega$.}. Its positivity leads to inequalities that are more specific to superfluids, like e.g. the positivity of both the normal and the superfluid mass densities. 
\item The ``causality term'' (i.e. all pieces that contain $V^\nu$) ensures that the system remains stable also in reference frames that are highly boosted relative to the environment. Including this term in the quadratic form causes all sound speeds to be subluminal \cite{GavassinoCausality2021,GavassinoUniversalityI2023,GavassinoSuperluminal2021,GavassinoBounds2023}. 
\end{itemize}
In the case of a single-component non-relativistic superfluid, the resulting inequalities coincide with those found in \cite{Andreev2004}. The real novelty here lies in the extension of these inequalities to superfluids with an arbitrary number of chemical components, and within full General Relativity.

In a forthcoming study, the inequalities \eqref{neutronstore}–\eqref{neutronstore2} will be used to derive upper bounds on the superfluid velocity in gapless neutron-star superfluids. Such thermodynamic bounds are particularly valuable, as Landau’s criterion does not apply in this context, potentially allowing the order parameter to remain finite well above the critical velocity \cite{AllardChamel2023PartI,AllardChamel2023PartII,AllardChamel2024EPJA,AllardChamel2024PRL,AllardChamel2025}. The study will also explore the resulting implications for neutron-star cooling and pulsar glitches.

We emphasize once more that our analysis is \textit{macroscopic}, as it is based on perturbations that are hydrodynamic in nature. Consequently, even if a superfluid state meets all the criteria outlined above, it could still be destabilized by microscopic processes involving transient, nonthermal quasiparticle configurations. For this reason, our findings (and similarly those of \cite{Andreev2004}) should be interpreted as necessary, but not sufficient, conditions for stability.

}




\section*{Acknowledgements}

This is partially supported by a Vanderbilt's Seeding Success Grant. I thank N. Chamel and V. Allard for bringing the problem to our attention, and for providing useful feedback.

\appendix








\section{Minimization of the grand-potential with pinned vortices}\label{pinnezzo}

When a superfluid is crossed by a lattice of quantum vortices, which are locally parallel to each other, one can adopt an effective coarse-grained description, where all hydrodynamic fields are averaged over many vortices \cite{langlois98,Gusakov:2016eom,GavassinoIordanskii2021}. The resulting effective theory (which is valid only at lengthscales that are much larger than the inter-vortex separation) is, to a first approximation, identical to the hydrodynamic theory provided in section \ref{section2}, with only one difference: the differential form $\mu_\nu$ (we drop the index $I$ for clarity) is no longer closed. Specifically, one can show that the coarse-grained vorticity $\varpi_{\nu \rho}=\nabla_{\nu}\mu_{\rho}-\nabla_{\rho}\mu_{\nu}$ is related to the density $\mathfrak{N}$ of vortices per unit area by the formula
\begin{equation}
\mathfrak{N}=\sqrt{\dfrac{\varpi^{\nu \rho}\varpi_{\nu \rho}}{2}}.
\end{equation}
Therefore, at these scales, $\nabla_\nu \mu_\rho \neq \nabla_\rho \mu_\nu$, and we can no longer write $\mu_\nu=\nabla_\nu \varphi$. Intuitively, this makes sense, since a vortex is a singularity of $\varphi$, and there are vortices crossing all fluid cells, meaning that $\varphi$ is ``everywhere singular''. The consequence is that, while equation \eqref{phioddottorne} is still true with vortices, the manipulations in \eqref{irrez} are no longer allowed. This raises the question: What are the local minima of $\Omega$ in the presence of a vortex lattice?

\textcolor{black}{The answer depends on the dynamics of the vortices. If the vortices are free to move, we must allow for an arbitrary vortex displacement $\dot{\varpi}_{\nu \rho}$ in our variation, and we are back to the analysis of \cite{GavassinoStabilityCarter2022}, which did not find any other minima besides full equilibrium. We have thus rediscovered, from thermodynamic reasoning alone, a well-known fact: when quantum vortices are free to move, every superfluid component is subject to an entropic force (the vortex-mediated mutual friction \cite{langlois98,Geo2020,GavassinoIordanskii2021}) which gradually damps all relative motions until the system comes to rest with respect to the environment. This force is generally taken to be the key driver of pulsar glitches.} 

Instead, suppose that the vortices are pinned to the environment, and that no viable fluctuation unpins them. Then, we must impose \cite{langlois98}
\begin{equation}
\beta_\star^\nu \varpi_{\nu \rho}=0 \, ,
\end{equation}
and Cartan's magic formula gives $(\mathfrak{L}_{\beta_\star}\varpi)_{\nu \rho}=0$. This means that $\varpi_{\nu \rho}$ is conserved, and it cannot change due to the perturbation, namely $\varpi_{\nu \rho}(\lambda)=\varpi_{\nu \rho}(0)$ (recall that we are assuming that no fluctuations can unpin the vortices). Therefore, we must set $\dot{\varpi}_{\nu \rho}=0$, which means that $\dot{\mu}_\nu$ is a closed differential form. Assuming that the domain covered by the superfluid in the coarse-grained description is simply-connected, we find that, although the relation $\mu_\nu=\nabla_\nu \varphi$ is \textit{not} valid, the relation $\dot{\mu}_\nu=\nabla_\nu \dot{\varphi}$ is valid, for some field $\dot{\varphi}$. Consequently, equation \eqref{thphidot} is again true, and all the manipulations that lead to Theorem \ref{theo1} still apply, completely unchanged. \textcolor{black}{We have thus rediscovered another well-known fact: The pinned regime is frictionless, as long as the superfluid phase is thermodynamically stable at the given relative velocities.}

\section{Comparison with the Andreev-Melnikovsky method}\label{comparisonwithandreev}

\citet{Andreev2004} imposed that the symmetric matrix $\partial p_a/\partial q^b$ be positive definite, with 
\begin{equation}
\begin{split}
p_a={}& (T, \, \Tilde{\mu}{+}v_S^k v_{Sk}/2{-}v_S^k u_k, \, J_k{-}\rho u_k, \, u_k) \, , \\
q^b={}& (s, \, \rho, \, v_{S}^k, \, J^k) \, , \\
\end{split}
\end{equation}
where $v_S^k=u^k+V_S^k$ is the superfluid velocity in the laboratory frame, $\Tilde{\mu}=\Bar{\mu} -\frac{1}{2}V_S^k V_{Sk}$ in the chemical potential in the superfluid rest frame, and $J^k= \rho u^k +\rho_S V_S^k$ is the mass flux. But if $\partial p_a/\partial q^b$ is positive definite, then the respective quadratic form
\begin{equation}
\mathcal{E}=\dfrac{1}{2} \delta q^a \, \dfrac{\partial p_a}{\partial q^b} \, \delta q^b = \dfrac{1}{2} \delta q^a \delta p_a
\end{equation}
is positive for all choices of $\delta q^a$. Our goal now is to show that the quadratic form above coincides with \eqref{TheMainENewt33}, meaning that the associated stability conditions are the same. This task is accomplished with simple algebraic manipulations:
\begin{equation}
\begin{split}
2\mathcal{E} ={}& \delta s \, \delta T +\delta \rho \, \delta \bigg(\Tilde{\mu}{+}\dfrac{v_S^k v_{Sk}}{2}{-}v_S^k u_k\bigg) +\delta v_{S}^k \delta (J_k{-}\rho u_k) \, +\delta J^k \delta u_k \\
={}& \delta s \, \delta T +\delta \rho \, \delta \bigg(\Bar{\mu}{-}\dfrac{u^k u_{k}}{2}\bigg) +\delta( u^k {+} V_{S}^k) \delta (\rho_S V_{Sk})  +\delta (\rho u^k {+}\rho_S V_S^k) \delta u_k \\
={}& \delta s \, \delta T +\delta \rho \, \delta \Bar{\mu}+ \rho \delta u^k \delta u_k +\delta V_{S}^k \delta (\rho_S V_{Sk})+2 \delta u^k\delta(\rho_S V_{Sk}) \, . \\
\end{split}
\end{equation}

\bibliography{Biblio}

\label{lastpage}

\end{document}